# Quantum Teleportation for Control of Dynamical Systems and Autonomy

Farbod Khoshnoud
Electromechanical Engineering Technology Department, College of Engineering, California State Polytechnic University, Pomona, CA 91768, USA.
Center for Autonomous Systems and Technologies, Department of Aerospace Engineering, California Institute of Technology 1200 E California Blvd, Pasadena, CA 91106, USA.

Lucas Lamata
Atomic, Molecular and Nuclear Physics Department, University of Seville, 41080 Sevilla, Spain.

Clarence W. de Silva
Department of Mechanical Engineering, University of British Columbia, Vancouver, BC V6T 1Z4, Canada.

Marco B. Quadrelli
Jet Propulsion Laboratory, California Institute of Technology, 4800 Oak Grove Drive, Pasadena, CA 91109, USA.

**Abstract**

The application of Quantum Teleportation for control of classical dynamic systems and autonomy is proposed in this paper. Quantum teleportation is an intrinsically quantum phenomenon, and was first introduced by teleporting an unknown quantum state via dual classical and Einstein-Podolsky-Rosen channels in 1993. In this paper, we consider the possibility of applying this quantum technique to autonomous mobile classical platforms for control and autonomy purposes for the first time in this research. First, a review of how Quantum Entanglement and Quantum Cryptography can be integrated into macroscopic mechanical systems for controls and autonomy applications is presented, as well as how quantum teleportation concepts may be applied to the classical domain. In quantum teleportation, an entangled pair of photons which are correlated in their polarizations are generated and sent to two autonomous platforms, which we call the Alice Robot and the Bob Robot. Alice has been given a quantum system, i.e. a photon, prepared in an unknown state, in addition to receiving an entangled photon. Alice measures the state of her entangled photon and her unknown state jointly and sends the information through a classical channel to Bob. Although Alice's original unknown state is collapsed in the process of measuring the state of the entangled photon (due to the quantum non-cloning phenomenon), Bob can construct an accurate replica of Alice's state by applying a unitary operator. This paper, and the previous investigations of the applications of hybrid classical-quantum capabilities in control of dynamical systems, are aimed to promote the adoption of quantum capabilities and its advantages to the classical domain particularly for autonomy and control of autonomous classical systems.

1. Introduction

Quantum teleportation is a fundamental quantum concept that allows one to distribute quantum states between distant parties, without measuring or having information about them. In future networks of robots with quantum processing, quantum communication, and quantum sensing capabilities, this quantum primitive, quantum teleportation, might enable a better communication between robots. This will allow one, for example, to transfer the outcome of a quantum processing from one robot to a distant robot, for further quantum processing starting with this quantum state.

Quantum Teleportation by teleporting an unknown quantum state via dual classical and Einstein-Podolsky-Rosen channels was introduced in 1993 [1], which has been demonstrated experimentally [2] and with deterministic approaches (e.g., [3]-[4]). Measurement of the Bell operator and quantum teleportation was introduced in 1995 [5]. Since then, various efforts in development of Quantum Teleportation have been carried out including: Efficient Teleportation Between Remote Single-Atom Quantum Memories [6], Gain tuning for continuous-variable quantum teleportation of discrete-variable states [7], Unconditional Quantum Teleportation [8], Complete quantum teleportation using nuclear magnetic resonance [9], probabilistic resumable quantum teleportation of a two-qubit entangled state [10], Quantum Teleportation Between Discrete and Continuous Encodings of an Optical Qubit [11], Quantum teleportation over the Swisscom telecommunication network [12], and Quantum teleportation-based state transfer of photon polarization into a carbon spin in diamond [13].

The organization of the paper is as follows. A review of Quantum Multibody Dynamics, Controls, Robotics and Autonomy ([14]-[17]) is given. In this review, quantum entanglement (Section 2.1) and quantum cryptography (Section 2.2) are used for hybrid classical-quantum control of classical multi-agent autonomous systems. In Section 2.3, the concept of Quantum Teleportation in conjunction with application to dynamical systems for autonomy is introduced.

## 2. Quantum multibody dynamics

Quantum Multibody Dynamics is referred to as the subject of applying quantum physical phenomena (such as quantum entanglement and superposition) integrated with control of a distributed classical dynamical system (such as multiple robots and autonomous systems), and analysing the resulting behaviour of the classical dynamic system when the quantum phenomenon is leveraged for control or communication purposes. Examples include the application of quantum entanglement and quantum cryptography protocols to control of robotic systems presented below, which is then further extended to the application of Quantum Teleportation for control and autonomy purposes in this paper as follows. A review on our proposal for how quantum entanglement and quantum cryptography can be integrated into physical mechanical systems for control and autonomy applications ([14]-[17]) is given in Section 2.1 and Section 2.2.

### 2.1 Quantum entanglement for dynamic systems

An experimental setup for quantum entanglement is shown in Figure 1. The proposed procedure of using quantum entangled photons for applications in control of autonomous platforms (Figure 1 to Figure 3), is presented as follows ([14]-[17]), as a hybrid quantum-classical process:

Quantum part:
- Single photons (pump photons) of 405 nm wavelengths, are generated by a laser diode source.
- Spontaneous parametric down-conversion (SPDC) process is carried out using a nonlinear crystal, beta-barium borate (BBO), to split the 405 nm photon beams into correlated entangled pairs of 810 nm wavelength photons. The entangled photon pairs thus generated that are orthonormal in their polarizations.
- The entangled photons are sent from the BBO crystal to two beamsplitters (BS) that pass the photons with horizontal polarization and reflect the photons with vertical polarization.
- Four Single Photon Counter (SPC) modules are placed as in Figure 1 to Figure 3, where each SPC counts the number of photons reaching them, and keeps track of the corresponding horizontal and polarizations (as they pass or are reflected by the beamsplitter, respectively).
- There are two SPC modules on each autonomous platform. Therefore, each of the autonomous platforms receives an entangled photon as explained above. This is interpreted as entangling the autonomous platforms (i.e., robots are sharing entangled photon pairs) by corresponding entangled photons that are detected by the SPC modules.

- A Coincidence Counter module records the time at which the photons are reaching the SPC modules. If the time that a photon pair reaches two SPCs (on two robots), regardless of the polarization of the photons, is within a small enough time window (e.g., less than 10 ns), then the photon pair are considered as correlated entangled pairs.

Classical part:

- The horizontal and vertical polarizations that are received by the SPC modules are converted to 0 and 1 digital signals, respectively, to be used for control of the robotic platforms.
- The corresponding digital signals that are obtained from the entangled photon polarizations are sent to digital microcontrollers onboard.
- Desired control tasks, such as motion commands sent to servomotors onboard, are defined correspondingly to the digital signals received.
- The polarizers in between the BBO and the beamsplitters (Figure 1 and Figure 2) is used as controller tools. They provide the capability of controlling the polarization of the photons as they pass through them, which alter the polarization of the entangled photons (e.g., horizontal and vertical polarizations).

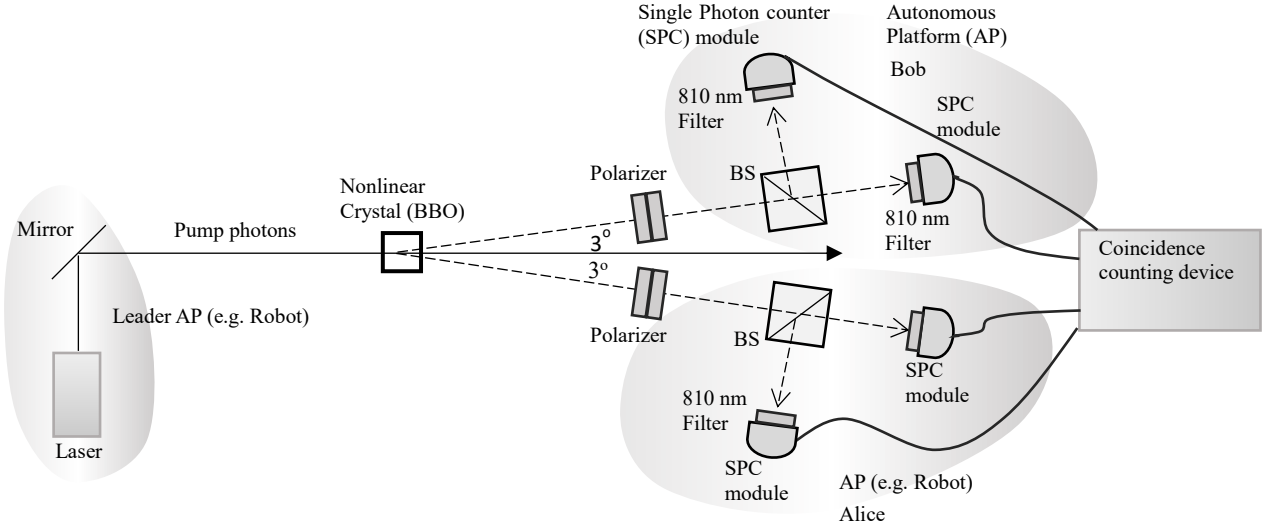

Figure 1. An experimental setup for quantum entanglement of autonomous platforms.

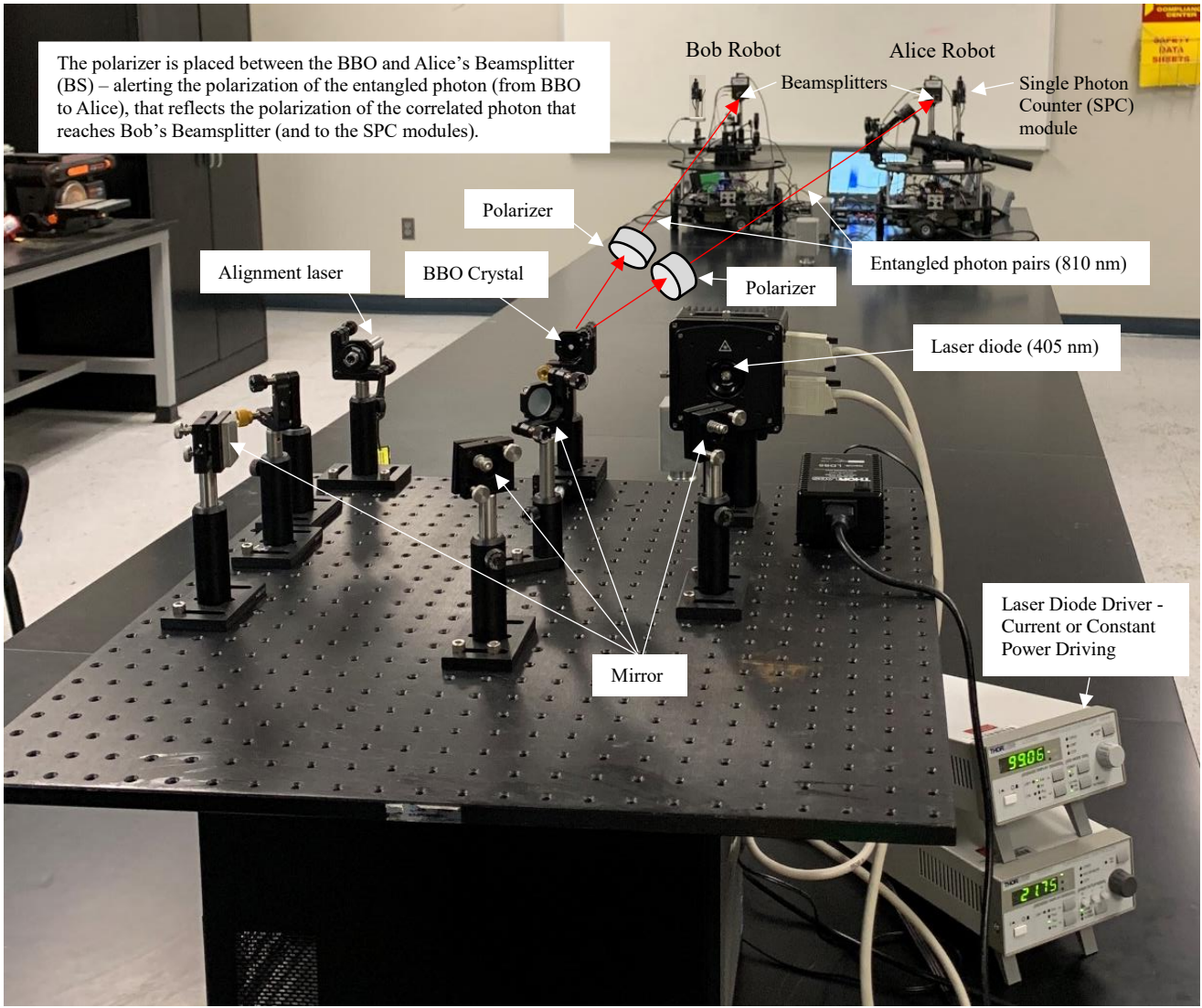

Figure 2. Quantum entanglement of autonomous platforms.

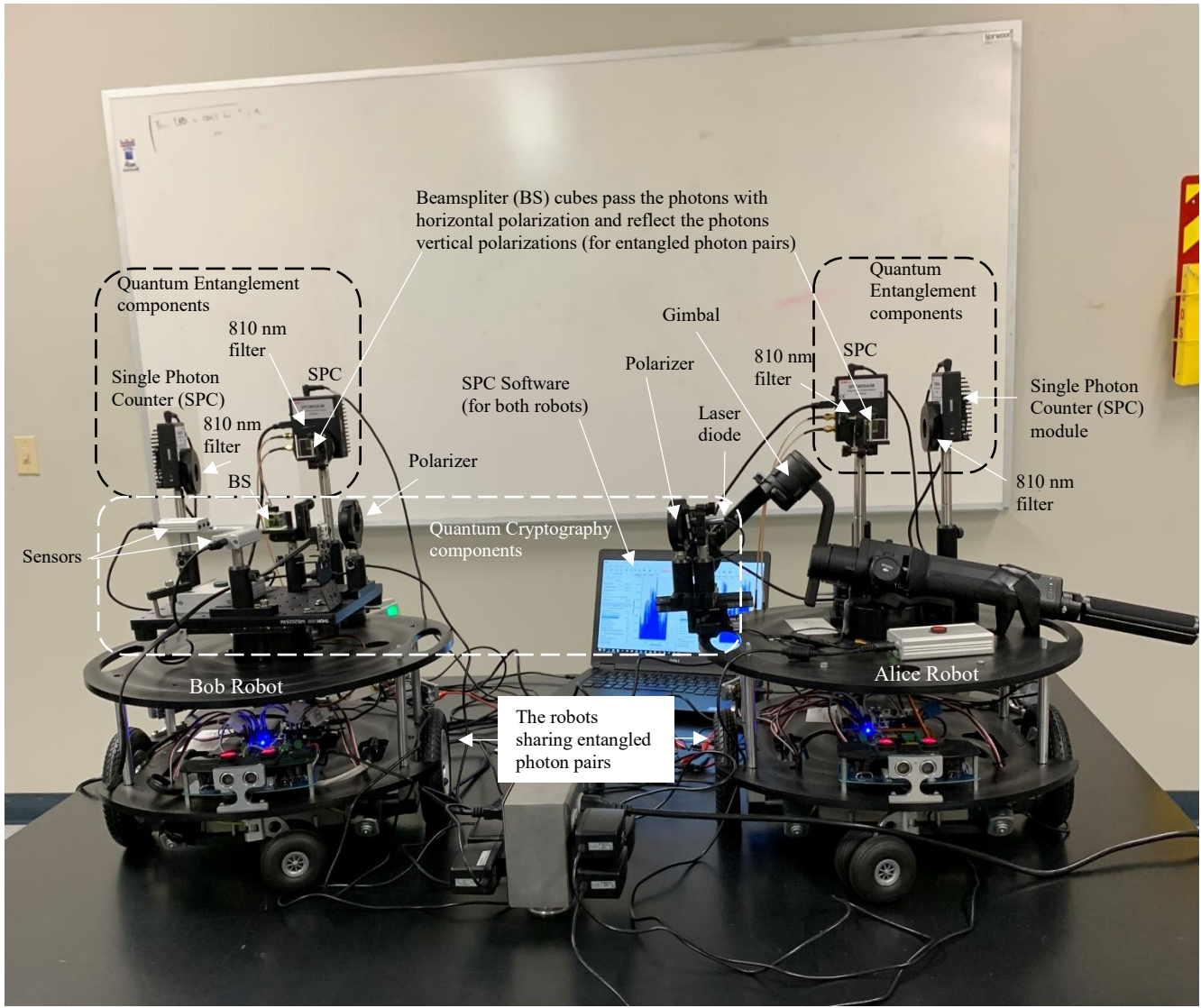

Figure 3. Autonomous platforms sharing quantum entangled photon pairs (the Alice Robot and the Bob Robot).

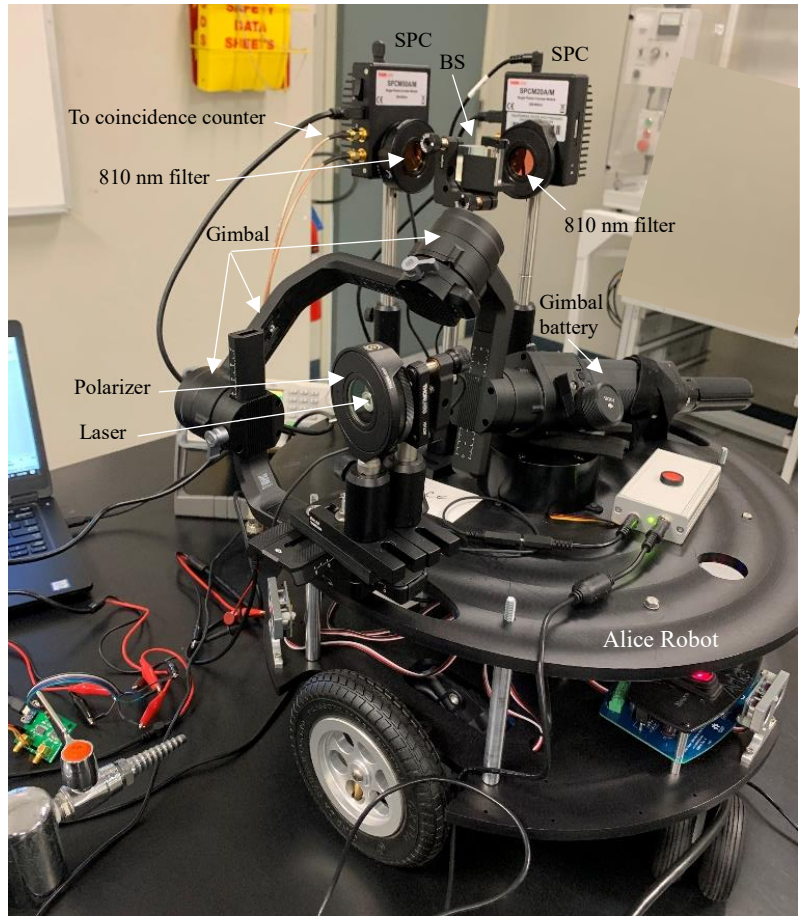

Figure 4. The Alice Robot.

## 2.2 Quantum cryptography for dynamic systems

The quantum cryptography setup for control of autonomous platforms (e.g., robots) is illustrated in Figure 5. This setup is installed on the same mobile robots in Figure 1 to Figure 4. Only the quantum cryptography system components are shown in Figure 5 (there is no quantum entanglement system component in Figure 5) for clear illustration of the setup. The quantum cryptography technique can be used to encrypt and transfer control commands from the Alice Robot to the Bob Robot (Figure 1 to Figure 5). The quantum cryptography communication technique can be used on its own for control of autonomous systems. In the case of using quantum cryptography in conjunction with quantum entanglement, one technique is to first entangle the robots by the quantum entanglement technique (Section 2.1), and then use the entanglement as a trigger to start the transfer the control commands from the Alice Robot to the Bob Robot.

The Quantum Cryptography process is as follows.

Quantum part:

- A single photon is sent from the Alice Robot (Figure 4).
- A polarizer placed on the Alice Robot is used to control the polarization of the photon as $|-45°\rangle$, $|0°\rangle$, $|45°\rangle$, and $|90°\rangle$.
- A polarizer placed on the Bob Robot has additional control on the photon polarization with $|0°\rangle$, $|45°\rangle$ orientations.
- After passing through the two polarizers, the photon reaches the beamsplitter which allows the photons with horizontal polarizations to pass, and the photons with vertical polarizations to reflect.
- There is a dedicated sensor for each direction of the photon, which either passes or is reflected by the beamsplitter.

Classical part:
- The sensor that is dedicated for detecting the horizontally polarized photons sends a digital 0 signal to a digital microcontroller every time that receives a photon. The sensor that receives vertical polarized photons sends digital 1 signal to the microcontroller.
- Desired control commands are defined based on these digital signals for application of autonomy and robotic tasks.

The detailed discussions of the quantum cryptography protocols and the detection of the eavesdropper are presented in references [14]-[17]. Autonomous platforms that are considered in this research (in integrating quantum technologies with mechanical systems) include any stationary or mobile system such as the ground robots in Figure 2 to Figure 4, or aerial systems such as the drones in Figure 7 and Figure 8.

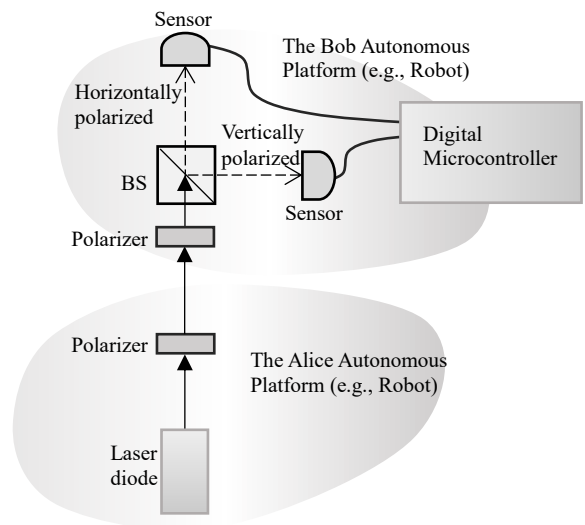

Figure 5. Quantum cryptography.

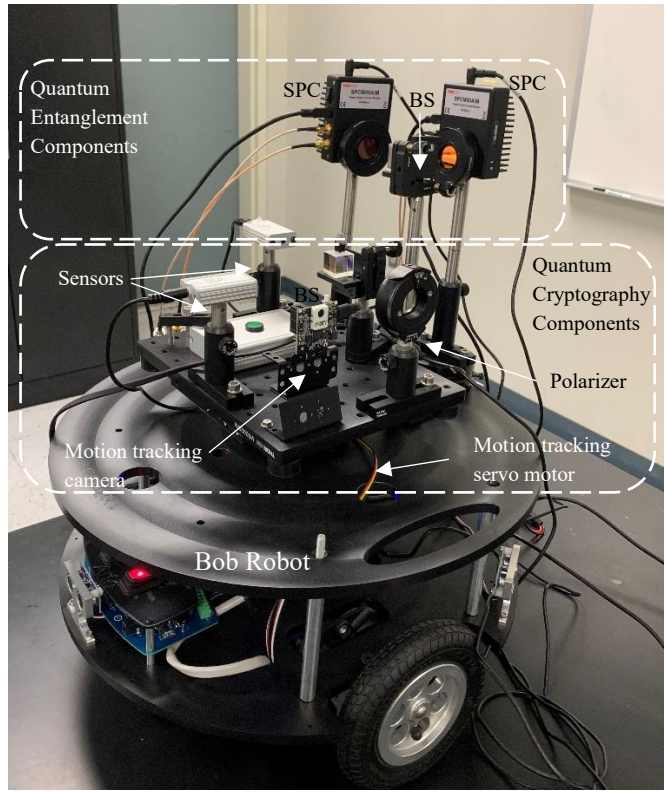

Figure 6. The Bob Robot.

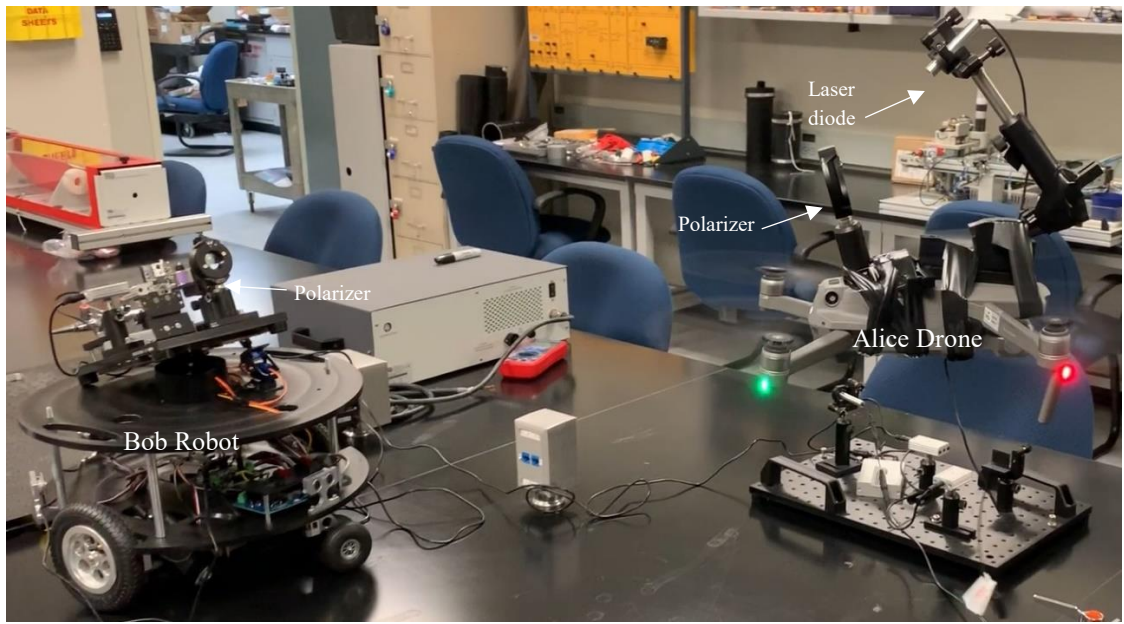

Figure 7. The Bob Ground Robot and the Alice Drone.

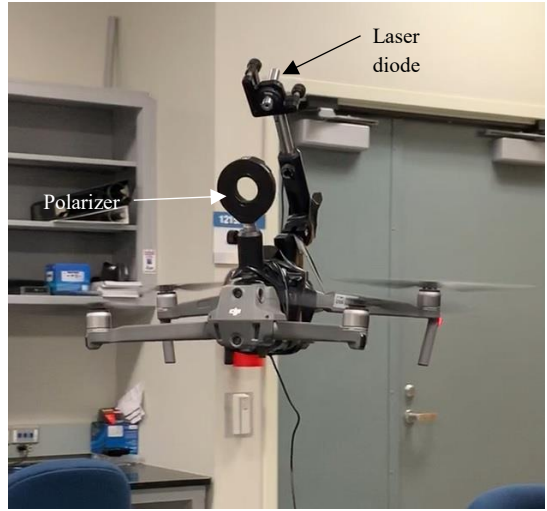

Figure 8. The Alice Drone.

**2.3 Quantum teleportation for dynamic systems**

The application of Quantum Teleportation for quantum control of classical dynamic systems and autonomy is proposed here. The Quantum Teleportation technique presented in this section is based on 'teleporting an unknown quantum state via dual classical and Einstein-Podolsky-Rosen (EPR) channels' [1]. In a classical system a bit can be copied and transferred (or cloned). By the "No-Cloning Theorem" [18], a quantum state as a quantum bit, or qubit, unlike the classical bit, can not be copied or cloned. Transfer of a qubit from one position (Alice) to another (Bob) is proposed to be carried out by a Quantum Teleportation technique [1]. Quantum Teleportation is applicable to quantum information, quantum cryptography, and quantum control areas. If a qubit is physically (classically) transported, for instance from A to B, the required quantum information can be lost. This is due to the sometimes short coherence times of the quantum states. The Quantum Entanglement phenomenon allows Quantum Teleportation under the assumption that strong correlation between quanta can be maintained.

We propose to apply quantum teleportation of dual classical and Einstein-Podolsky-Rosen (EPR) channels technique to mechanical systems, such as autonomous mobile platforms, for control and autonomy purposes here for the first time. In quantum teleportation, an entangled pair of photons which are correlated in their polarizations are generated and sent to two autonomous platforms, which we call Alice Robot and Bob Robot. Alice has been given a quantum system, i.e. a photon, prepared in an unknown state. Alice also receives one of the entangled photons. Alice measures the state of her entangled photon and sends the information through a classical channel to Bob. Although Alice's original unknown state is collapsed in the process of measurement, due to quantum non-cloning phenomenon, Bob can construct an accurate replica of Alice's state by applying a unitary operator.

This paper and the previous investigations of the applications of quantum capabilities in control of dynamical systems are aimed to promote the adoption of unmatched quantum capabilities and advantages in the classical domain particularly for autonomy and control of autonomous systems. In this context, the basic scheme of quantum teleportation is as in Figure 1 and Figure 2. In Quantum Teleportation, the goal is to transfer a quantum state $|\phi_1\rangle$ from the Alice Robot to the Bob Robot. The quantum teleportation process can be notionally described as follows.

- Alice generates the quantum state $|\phi_1\rangle$ to transfer to Bob, where $|\phi_1\rangle = a|0\rangle_1 + b|1\rangle_1$, with $|a|^2 + |b|^2 = 1$.
- Quantum entangled states are sent from the BBO (Figure 1 and Figure 2) to Alice and Bob by the SPDC process. Using Bell basis, the basis of a pair of entangled photons (e.g., a two-qubit-system), which are correlated by means of their polarizations, is given by

$$|\Psi_{23}^{\pm}\rangle = \frac{1}{\sqrt{2}}(|0\rangle_2|1\rangle_3 \pm |1\rangle_2|0\rangle_3)$$

$$|\Phi_{23}^{\pm}\rangle = \frac{1}{\sqrt{2}}(|0\rangle_2|0\rangle_3 \pm |1\rangle_2|1\rangle_3)$$

- Alice and Bob receive the entangled photons.
- Alice makes a complete measurement of the entangled photon (that receives), and the state $|\phi_1\rangle$. Thus, the measurement of the state $|\phi_1\rangle$ and the entangled state $|\Psi_{23}^{-}\rangle$ can be represented as

$$|\Psi_{123}\rangle = \frac{a}{\sqrt{2}}(|0\rangle_1|0\rangle_2|1\rangle_3 - |0\rangle_1|1\rangle_2|0\rangle_3) + \frac{b}{\sqrt{2}}(|1\rangle_1|0\rangle_2|1\rangle_3 - |1\rangle_1|1\rangle_2|0\rangle_3),$$

- After Alice's measurement, Bob's particle (with indices 3) will have been projected into one of the four pure states superposed, with equal possibilities. Bob's particle 3 can be presented by the products of the states in terms of the Bell operator basis as

$$|\Psi_{123}\rangle = \frac{1}{2}[|\Psi_{12}^{-}\rangle(-a|0\rangle_3 - b|1\rangle_3) + |\Psi_{12}^{+}\rangle(-a|0\rangle_3 + b|1\rangle_3) + |\Phi_{12}^{-}\rangle(a|1\rangle_3 + b|0\rangle_3) + |\Phi_{12}^{+}\rangle(a|1\rangle_3 - b|0\rangle_3)]$$

- One of the above Bob's entangled states (one of the four possibilities) is related to the state, $|\phi_1\rangle$, that Alice is to teleport to Bob. Bob is required to produce a replica of Alice's state.
- Alice's measurement is transmitted as classical information through a classical channel to Bob.
- Bob uses the transmitted information from Alice and applies a set of corresponding unitary transformations on his EPR, according to the Alice's transmitted state. The unitary transformations are $\hat{U}_1 = -\begin{bmatrix} 1 & 0 \\ 0 & 1 \end{bmatrix}$, $\hat{U}_2 = \begin{bmatrix} -1 & 0 \\ 0 & 1 \end{bmatrix}$, $\hat{U}_3 = \begin{bmatrix} 0 & 1 \\ 1 & 0 \end{bmatrix}$, and $\hat{U}_4 = \begin{bmatrix} 0 & -1 \\ 1 & 0 \end{bmatrix}$.

- This transformation brings Bob's particle to the state of Alice's particle 1 ($|\phi_1\rangle$), and the teleportation process is complete.

On the application of quantum teleportation in the context of robotics and autonomy schemes, the Alice Robot teleports a state $|\phi_1\rangle$ to the Bob Robot. Bob converts the polarization information received from Alice into corresponding 0 and 1 digital information, which is then processed by on-board microcontrollers for performing predefined robotic and autonomy tasks.

By accessing quantum computers in future, the on-board classical computation using microcontrollers may be instead realized by quantum processors. In fact, the application of future quantum computers in a network of multi-agent dynamic systems, is only logical if quantum-based techniques such as entanglement and teleportation are employed, rather than any classical wireless communication protocol in the communication network. In a network of autonomous platforms where multiple robotic agents containing quantum processors are communicating with each other, using any classical communication technique may actually defeat the advantages of quantum-enhanced protocols.

**Conclusion**

We introduced the integration of quantum mechanical phenomena, such as quantum entanglement, into classical mechanical systems, such as mobile autonomous platforms, as a hybrid classical-quantum system. In particular, the concept of quantum teleportation by teleporting a quantum state via dual classical and Einstein-Podolsky-Rosen channels in the context of the control of dynamical systems and autonomy was proposed. A review of the applications of quantum entanglement and quantum cryptography in developing quantum-enhanced networks of robotic systems was presented. A proposed procedure of how quantum technologies could be brought into the domain of classical mechanical systems by employing quantum entanglement, cryptography and teleportation was described. The research outlined in this paper serves as a first step towards the application of the advantages of quantum techniques in the physical domain of macroscopic dynamic systems. Furthermore, this investigation aims to promote future attempts at exploring the interdisciplinary interface of quantum mechanics and classical system autonomy schemes, by pushing the engineering boundaries beyond any existing classical technique. Using on-board quantum processors, instead of classical microcontrollers, is proposed as one future direction of this research.

**Acknowledgement**

Lucas Lamata acknowledges the funding from PGC2018-095113-B-I00, PID2019-104002GB-C21, and PID2019-104002GB-C22 (MCIU/AEI/FEDER, UE). Government sponsorship acknowledged. Dr. Quadrelli's contribution was carried out at the Jet Propulsion Laboratory, California Institute of Technology, under a contract with the National Aeronautics and Space Administration.

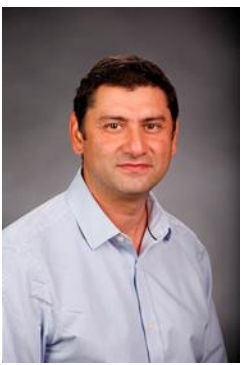

Farbod Khoshnoud, PhD, CEng, PGCE, HEA Fellow, is a faculty member in the college of engineering at California State Polytechnic University, Pomona, and a visiting associate in the Center for Autonomous Systems and Technologies, in the department of Aerospace Engineering at California Institute of Technology. His current research areas include Self-powered and Bio-inspired Dynamic Systems; Quantum Multibody Dynamics, Robotics, Controls and Autonomy, by experimental Quantum Entanglement, and Quantum Cryptography; and theoretical Quantum Control techniques. He was a research affiliate at NASA's Jet Propulsion Laboratory, Caltech in 2019, an Associate Professor of Mechanical Engineering at California State University, 2016-18, a visiting Associate Professor in the Department of Mechanical Engineering at the University of British Columbia (UBC) in 2017, a Lecturer in the Department of Mechanical Engineering at Brunel University London, 2014-16, a senior lecturer and lecturer at the University of Hertfordshire, 2011-14, a visiting scientist and postdoctoral researcher in the Department of Mechanical Engineering at UBC, 2007-11, a visiting researcher at California Institute of Technology, 2009-11, a Postdoctoral Research Fellow in the Department of Civil Engineering at UBC, 2005-2007. He received his Ph.D. from Brunel University in 2005. He is an associate editor of the Journal of Mechatronic Systems and Control.

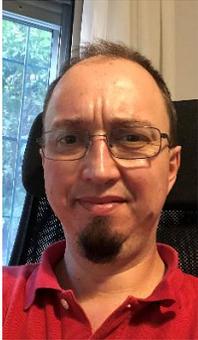

Prof. Lucas Lamata is an Associate Professor (Profesor Titular de Universidad) of Theoretical Physics at the Departamento de Física Atómica, Molecular y Nuclear, Facultad de Física, Universidad de Sevilla, Spain. His research up to now has been focused on quantum optics and quantum information, including pioneering proposals for quantum simulations of relativistic quantum mechanics, fermionic systems, and spin models, with trapped ions and superconducting circuits. He also analyzes the possibility of combining artificial intelligence and machine learning protocols with quantum devices. Before working in Sevilla, HeI was a Staff Researcher (Investigador Doctor Permanente) at the University of the Basque Country, Bilbao, Spain (UPV/EHU), leading the Quantum Artificial Intelligence Team, a research group inside the QUTIS group of Prof. Enrique Solano at UPV/EHU. Before that, he was a Humboldt Fellow and a Max Planck postdoctoral fellow for 3 and a half years at the Max Planck Institute for Quantum Optics in Garching, Germany, working in Prof. Ignacio Cirac Group. Previously, he carried out his PhD at CSIC, Madrid, and Universidad Autónoma de Madrid (UAM), with an FPU predoctoral fellowship, supervised by Prof. Juan León. He has more than 100 articles, among published and submitted, in international refereed journals, including: 1 Nature, 1 Reviews of Modern Physics, 1 Advances in Physics: X, 3 Nature Communications, 2 Physical Review X, and 19 Physical Review Letters, two of them Editor's Suggestion. His h-index according to Google Scholar is of 35, with more than 4400 citations.

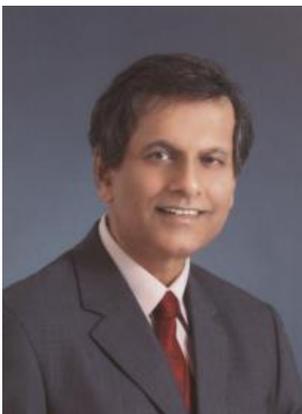

Clarence W. de Silva has been a Professor of Mechanical Engineering at the University of British Columbia, Vancouver, Canada since 1988. He received Ph.D. degrees from Massachusetts Institute of Technology and University of Cambridge, U.K., honorary D.Eng. degree from University of Waterloo, Canada, and the higher doctorate (ScD) from University of Cambridge. He is a Fellow of: IEEE, ASME, Canadian Academy of Engineering, and Royal Society of Canada. Also, he has been a Senior Canada Research Chair, NSERC-BC Packers Chair in Industrial Automation, Mobil Endowed Chair, Lilly Fellow, Senior Fulbright Fellow, Killam fellow, Erskine Fellow, Professorial Fellow, Faculty Fellow, Distinguished Visiting Fellow of Royal Academy of Engineering, UK, and a Peter Wall Scholar. He has authored 25 books and over 550 papers, approximately half of which are in journals. His recent books published by Taylor & Francis/CRC are: Modeling of Dynamic Systems—with Engineering Applications (2018); Sensor Systems (2017); Senors and Actuators—Engineering System Instrumentation, 2nd edition (2016); Mechanics of Materials (2014); Mechatronics—A Foundation Course (2010); Modeling and Control of Engineering Systems (2009); VIBRATION—Fundamentals and Practice, 2nd Ed. (2007); by Addison Wesley: Soft Computing and Intelligent Systems Design—Theory, Tools, and Applications (with Karray, 2004), and by Springer: Force and Position Control of Mechatronic Systems—Design and Applications in Medical Devices (with Lee, Liang and Tan, 2020).

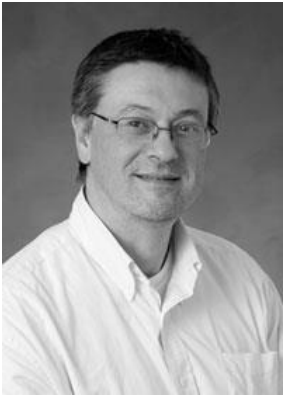Dr. Quadrelli is a Principal Member of the Technical Staff, and the supervisor of the Robotics Modeling and Simulation Group at JPL. He has a degree in Mechanical Engineering from Padova (Italy), an M.S. in Aeronautics and Astronautics from MIT, and a PhD in Aerospace Engineering from Georgia Tech. After joining NASA JPL in 1997 he has contributed to a number of flight projects including the Cassini-Huygens Probe, Deep Space One, the Mars Aerobot Test Program, the Mars Exploration Rovers, the Space Interferometry Mission, the Autonomous Rendezvous Experiment, and the Mars Science Laboratory, among others.  He has been the Attitude Control lead of the Jupiter Icy Moons Orbiter Project, and the Integrated Modeling Task Manager for the Laser Interferometer Space Antenna. He has led or participated in several independent research and development projects in the areas of computational micromechanics, dynamics and control of tethered space systems, formation flying, inflatable apertures, hypersonic entry, precision landing, flexible multibody dynamics, guidance, navigation and control of spacecraft swarms, terra-mechanics, and precision pointing for optical systems. He is an Associate Fellow of the American Institute of Aeronautics and Astronautics, a NASA Institute of Advanced Concepts Fellow, and a Caltech/Keck Institute for Space Studies Fellow.